\documentclass{llncs}
\usepackage{color}
\usepackage{hyperref}
\usepackage{verbatim}
\usepackage{xspace}
\usepackage{latexsym}
\usepackage{graphicx}
\usepackage{times}

\hypersetup{
    pdftitle={Learning Better Context Characterizations: An Intelligent Information Retrieval Approach},
}
\newcommand{\step}{\mathrm s}
\newcommand{\M}{{\bf H}}
\newcommand{\simi}{\sigma}
\newcommand{\nsimilarity}{\sigma^{N}}

\newcommand{\discD}{\delta}
\newcommand{\descD}{\lambda}
\newcommand{\descT}{\Lambda}
\newcommand{\discT}{\Delta}

\begin{document}

\title{Learning Better Context Characterizations:\\
 An Intelligent Information Retrieval Approach\thanks{This research
work is supported by Agencia Nacional de Promoci\'on Cient\'{\i}fica
y Tecnol\'ogica (PICT 2005 Nro. 32373) and Universidad Nacional del
Sur (PGI 24/ZN13).}}

\author{Carlos M. Lorenzetti \ \ \ \ \ \ \ Ana G. Maguitman} \institute{
Grupo de Investigaci\'on en Recuperaci\'on de Informaci\'on y
Gesti\'on del
Conocimiento\\
LIDIA - Laboratorio de Investigaci\'{o}n y Desarrollo en Inteligencia Artificial \\
    Departamento de Ciencias e Ingenier\'{\i}a de la Computaci\'{o}n \\
    Universidad Nacional del Sur,
    Av. Alem 1253, (B8000CPB) Bah\'{\i}a Blanca, Argentina\\
    CONICET - Consejo Nacional de Investigaciones Cient\'{\i}ficas y
    T\'ecnicas\\
    phone:  54-291-4595135 \ \  fax: 54-291-4595136\\
   e-mail: \texttt{\{cml,agm\}@cs.uns.edu.ar}
   }

\maketitle \thispagestyle{empty}

\begin{abstract}

This paper proposes an incremental method that can be used by an
intelligent system to learn better descriptions of a thematic
context. The method starts with a small number of terms selected
from a simple description of the topic under analysis and uses this
description as the initial search context. Using these terms, a set
of queries are built and submitted to a search engine. New documents
and terms are used to refine the learned vocabulary. Evaluations
performed on a large number of topics indicate that the learned
vocabulary is much more effective than the original one at the time
of constructing queries to retrieve relevant material.
\end{abstract}

\section{Introduction}
\label{sec:introduction}

 Today's search engine
interfaces are appropriate  when the user knows {\em what} to seek
and {\em how} to seek it. However, they are unable to reflect the
user context and therefore they are not smart enough to understand
the real user's needs. For several years researchers in the
Artificial Intelligent community have talked about the importance of
intelligent systems that cooperate with the user to facilitate a
number of computer mediated
task~\cite{licklider60manmachine,maes94agents}. More recently, the
problem of accessing relevant information through intelligent
systems has become a main research area. In order to implement
intelligent Information Retrieval (IR) systems some researchers have
proposed taking advantage of existing services to build more
powerful tools on top of
them~\cite{gordon88necessity,etzioni96moving}. Examples of systems
that apply this approach take advantage of major search engines to
perform intelligent context-based
search~\cite{balabanovic95adaptive,budzik01information,rhodes00justintime,maguitman05suggesting,ramirez06semantic}

The Web can be regarded as a rich repository of collective memory.
An intelligent system that incrementally searches this repository to
find material that is useful to the user's current needs can act as
a memory augmentation aid. By an association of similarities, this
aid can help users remember information, assure that areas relevant
to the current task have been considered,  and pursue new
directions.

Descriptions of a user's needs, however, are usually deficient
because they are typically based on the a priori knowledge of the
topic of interest. This knowledge might be insufficient to formulate
a good query, or more commonly, the vocabulary used by the user
might not be appropriate to target the request at the right kind of
material. In certain scenarios, attaining novelty and diversity may
be as important, or even more important, than attaining similarity.
For human-generated queries users frequently decide, based on
initial results, to refine subsequent queries. If contextual
information is available, part of the query formation and refinement
process can be automated.

This paper proposes a new technique for  incrementally learning a
better characterization of the user context. The work presented here
suggests and tests the following hypotheses: (1) the vocabulary
describing the initial context can be used to identify semantically
related documents and terms, but (2) the terms describing the
initial context are not necessarily the most appropriate ones to
generate search queries, and (3) the characterization of the search
context can be incrementally improved by a semi-supervised learning
algorithm.

Our algorithm is based on the dynamic extraction of topic
descriptors and discriminators, as first introduced in
\cite{maguitman04dynamic}. The main contribution of this paper is
the proposal of a new mechanism for learning rich vocabularies
associated with a thematic context. The learned vocabulary provides
an improved characterization of the topic of interest in the sense
that it allows to better identify topically relevant material. The
effectiveness of our proposal is assessed by carrying out a
comprehensive evaluation on a large collection of human-generated
topic descriptions.

\section{Context Characterizations}
\label{sec:contextual-search} \vspace{-0.3cm}

For many computer-mediated tasks, the user context provides a rich
set of terms that can be exploited by intelligent systems to
generate queries and present related information to the user. Such
systems can be equipped with special monitoring capabilities,
designed to generate a model of the user context. The system will be
in charge of observing how the user interacts with different kinds
of computer utilities (such as email systems, browsers and text
editors) to characterize the user's information needs as a
collection of weighted terms. This requires a framework for learning
context-specific terms.

\subsection{The Different Role of Terms}
A central question addressed in our work is how to learn
context-specific terms based on the user current context and an open
collection of incrementally retrieved documents. In what follows, we
will assume that a user context is represented  as a set of  terms.
Consider for example a topic involving the {\em Java Virtual
Machine}. Context-specific terms may play different roles. For
example, the term {\em java} is a good descriptor of the topic for a
general audience. However, {\em java} is not a good discriminator
for that topic because it might also refer to the island in
Indonesia,  the java shark, a brand of Russian cigarettes or a
variety of coffee grown on the island of Java, among other
possibilities.\footnote{Wikipedia disambiguation page presents more
than 50 senses for the word {\em java}.}

If we reconsider the topic {\em Java Virtual Machine} we notice that
terms such as {\em jvm} and {\em jdk}---which stand for ``Java
Virtual Machine'' and ``Java Development Kit''---may not be good
descriptors of the topic for a general audience, but are effective
in bringing information that is relevant for our topic of interest
when presented in a query. Therefore, {\em jvm} and {\em jdk} are
good discriminators of that topic.

A natural question that arises in this scenario is how to identify
the terms that act as good descriptors and good discriminators of a
topic. In previous work
\cite{maguitman04dynamic,lorenzetti07intelligent} we have studied
and tested the following two hypotheses: \vspace{-0.2cm}
\begin{itemize}
\item Good topic descriptors can be found by looking for terms that occur
\underline{often} in documents related to the given topic.
\item Good topic discriminators can be found by looking for terms that occur
\underline{only} in documents related to the given topic.
\end{itemize}\vspace{-0.2cm}

Both topic descriptors and
discriminators are important as query terms. Because topic
descriptors occur often in relevant pages, using them as query terms
may improve recall. Similarly, good topic discriminators occur
primarily in relevant pages, and therefore using them as query terms
may improve precision.

\subsection{Computing Topic Descriptors and Topic Discriminators}

As a first approximation to compute descriptive and discriminating
power, we begin with a collection of $m$ documents and $n$ terms. As
a starting point we build an $m\times n$ matrix $\M$, such that
${\bf \M}[i,j]=k$ if $k$ is the number of occurrences of term $t_j$
in document $d_i$. In particular we can assume that one of the
documents (e.g., $d_0$) corresponds to the initial user context.

The matrix $\M$ allows us to formalize the notions of good
descriptors and good discriminators.  We define {\em descriptive
power of a term in a document} as a function
$\descD:\{d_0,\ldots,d_{m-1}\}\times \{t_0,\ldots,t_{n-1}\}
\rightarrow [0,1]$: {\small
\[
\descD(d_i,t_j) = \frac{\M[i,j]}
                     {\sqrt{\sum_{k=0}^{n-1} (\M[i,k])^2}}.
\]}
\noindent Note that $\descD$ can be regarded as a version of matrix
$\M$ normalized by row (i.e, by document).

If we adopt $\step(k)=1$ whenever $k>0$ and $\step(k)=0$ otherwise,
we can define  the {\em discriminating power of a term in a
document} as a function $\discD:\{t_0,\ldots,t_{n-1}\} \times
\{d_0,\ldots,d_{m-1}\} \rightarrow [0, 1]$:{\small
\[
\discD(t_i,d_j) = \frac{\step(\M[j,i])}
                     {\sqrt{\sum_{k=0}^{m-1} \step(\M[k,i])}}.
\]}

\noindent In this case $\discD$ can be regarded as a transposed
version of matrix $\M$ normalized by column (i.e, by term).

 Our current goal is to learn a better
characterization of the user needs. Therefore rather than extracting
descriptors and discriminators directly from the user context, we
want to extract them from {\em the topic} of the user context. This
requires an incremental method to characterize the topic of the user
context, which is done by identifying documents that are similar to
the user current context. Assume the user context and the retrieved
documents are represented as document vectors in term space. To
determine how similar two documents $d_i$ and $d_j$ are we adopt the
IR cosine similarity \cite{baeza-yates99modern}. This measure is
defined as a function $\simi: \{d_0,\ldots, d_{m-1}\} \times
\{d_0,\ldots, d_{m-1}\} \rightarrow [0,1]$:{\small
\[
\simi(d_i,d_j) = \sum_{k=0}^{n-1}[\descD(d_i,t_k)\cdot
\descD(d_j,t_k)].
\]
}

We formally define the {\em term descriptive power in the topic of a
document} as a function $\descT : \{d_0,\ldots, d_{m-1}\} \times
\{t_0, \ldots, t_{n-1}\} \to [0,1]$.  We set
                                       $\descT(d_i,t_j)=0$ if
$\sum_{\stackrel{k=0}{k\neq i}}^{m-1}
                                       \simi(d_i,d_k) = 0$.
                                       Otherwise we define $\descT(d_i,t_j)$
                                       as follows: {\small
\[
\descT(d_i,t_j) =
                            \frac{\sum_{\stackrel{k=0}{k\neq i}}^{m-1}
                                       [\simi(d_i,d_k)\cdot [\descD(d_k,t_j)]^2]}
                      {\sum_{\stackrel{k=0}{k\neq i}}^{m-1}
                                       \simi(d_i,d_k)}.
\]}
\noindent Thus, the descriptive power of a term $t_j$ in the topic
of a document $d_i$ is a measure of the quality of $t_j$ as a
descriptor of documents similar to $d_i$.

Analogously, we define the {\em discriminating power of a term in
the topic of a document} as a function
$\discT:\{t_0,\ldots,t_{n-1}\}\times
\{d_0,\ldots,d_{m-1}\}\rightarrow [0,1]$ calculated as
follows:{\small
\[
\begin{array} {rl}
\discT(t_i,d_j) = &  \sum_{\stackrel{k=0}{k\neq
j}}^{m-1}[[\discD(t_i,d_k)]^2 \cdot \simi(d_k,d_j)].
\end{array}
\]}
\noindent Thus the discriminating power of term $t_i$ in the topic
of document $d_j$ is an average of the similarity of $d_j$ to other
documents discriminated by $t_i$. For a worked example showing the
results of computing topic descriptors and discriminators
see~\cite{lorenzetti07intelligent}.

\section{An Algorithm for Context Enrichment through Vocabulary Leaps}
\label{sec:algorithm} \label{sec:context} \vspace{-0.3cm} Attempting
to find an optimal set of terms to characterize the user thematic
context gives rise to a combinatorial problem. This is not only
intractable but unreasonable from a pragmatic point of view.
Instead, we propose to apply an intelligent IR strategy to explore
and exploit potentially useful vocabularies. Assume the vocabulary
defines a landscape, where the initial context is a given region of
this landscape. In this scenario, exploitation means to thoroughly
explore a given set of terms in order to find local optima, i.e.,
the best descriptors and discriminators based on a given
characterization of the current context. Exploration, on the other
hand, refers to probe new regions of the landscape, which is
dynamically discovered by performing incremental search, in the hope
of finding either better descriptors or better discriminators and
therefore a better characterization of the thematic context.
\begin{figure}[!ht]
\centerline{\includegraphics[scale=0.45]{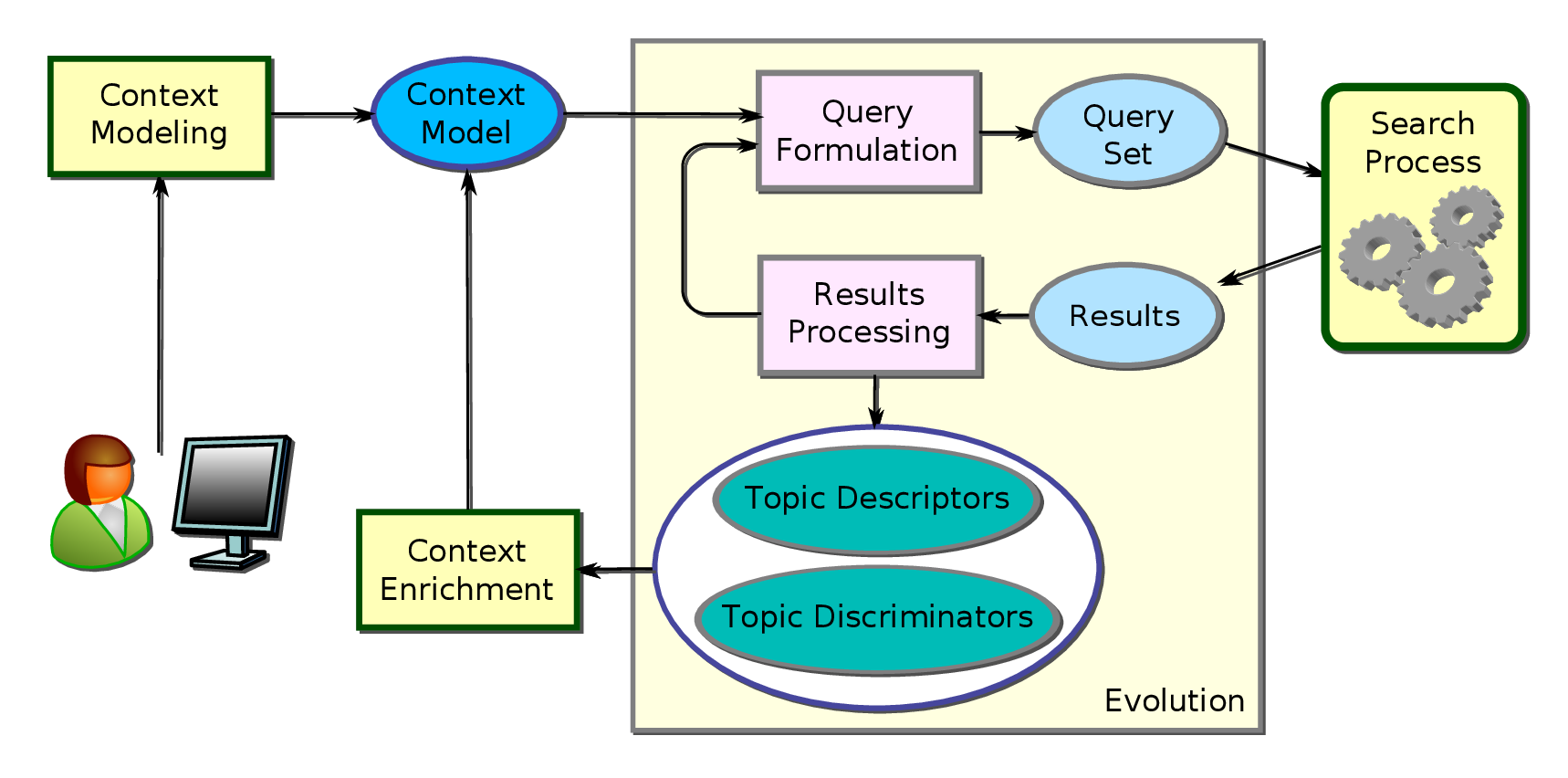}}
\caption{A schematic illustration of the proposed mechanism for
learning better context characterizations.} \label{fig:algdiagram}
\end{figure}

Many machine learning techniques that apply the
exploration-exploitation strategy (e.g., simulated annealing and
reinforcement learning) attempt to diversify (i.e., to explore)
during initial generation and to focus (i.e., to exploit) towards
the end. In our approach we take a different approach and propose an
algorithm that evolves topic descriptors and discriminators by
alternating the exploration and the exploitation of the vocabulary
landscape. We begin by exploiting the initial vocabulary by focusing
on the initial context. This vocabulary is used to iteratively form
queries that are submitted to a search engine. If after a certain
number of iterations there are no significant improvements on the
search results, our algorithm performs a phase change to explore new
potentially useful regions of the vocabulary landscape. A phase
change can be regarded as a vocabulary leap, which can be thought of
as a significant transformation (typically an improvement) of the
context characterization. A schematic illustration of the proposed
mechanism for learning better context characterization is shown in
figure~\ref{fig:algdiagram} and is summarized in the following
steps:{\small
\begin{enumerate}
    \item Let  $\mathcal{C}$ be the initial context description.
    \item Set $\mathcal{C}_0 = \mathcal{C}$.

    \item $i \leftarrow 0$, repeat
        \begin{enumerate}
        \item Start phase $\mathcal{P}_i$
            \item Set $\descT_{\mathcal{E}_0^i} = 0$ and $\discT_{\mathcal{E}_0^i} = 0$
        \item $j \leftarrow 1$, repeat
        \begin{enumerate}
            \item Start $\mathcal{P}_i$ evolution,
            $\mathcal{E}_j^i$.
            \item Set $\mathcal{Q}$ equal to some combination of context
            terms.\footnote{See section~\ref{sec:evaluation} for
            details on how the combination was implemented in our tests.}
            \item do Search with $\mathcal{Q}$.
            \item Make lists of topic descriptors and discriminators, $\descT'$ and $\discT'$, based on search results and
            $\mathcal{C}_i$.
            \item Update $\descT_{\mathcal{E}_j^i}$ and $\discT_{\mathcal{E}_j^i}:$
            \begin{itemize}
                \item $\{\descT_{\mathcal{E}_j^i}|\discT_{\mathcal{E}_j^i}\} = \alpha\{\descT_{\mathcal{E}_{j-1}^i}|\discT_{\mathcal{E}_{j-1}^i}\} +
                \beta\{\descT'|\discT'\}$.
            \end{itemize}
            \item Analyze the documents' similarity\footnote{In order to test
             our algorithm we used the measure of novelty-driven
            similarity defined in section~\ref{sec:evaluation} for
            reasons that will become obvious in that section.} to
            $\mathcal{C}_i$ every $u$ iterations:
            \begin{itemize}
                \item If there is a low variation ($\theta < \mu$), end $\mathcal{E}_j^i$. Return $\descT_{i}=\descT_{\mathcal{E}_j^i}$ and $\discT_{i}=\discT_{\mathcal{E}_j^i}$.
                \item If the process has run for at least $v$ iterations and there is a very low variation ($\theta <\nu$), end $\mathcal{P}_i$ and goto~\ref{step:end}.
            \end{itemize}
            \item $j \leftarrow j + 1$.
        \end{enumerate}
        \item   Update $\mathcal{C}_i$ with terms containing high  $\descT_{i}$ and
        $\discT_{i}$ values to obtain $\mathcal{C}_{i+1}$.
         \item Let
    $w_{\mathcal{C}_{i}}^{t_k}$ represent the weight of term $t_k$
    in context $\mathcal{C}_{i}$.
        \item Set the terms weights
                $w_{\mathcal{C}_{i+1}}^{t_k} = \gamma w_{\mathcal{C}_{i}}^{t_k} + \zeta w_{\descT_{i}}^{t_k} + \xi w_{\discT_{i}}^{t_k}$
        \item $i \leftarrow i + 1$.
    \end{enumerate}
    \item \label{step:end}End process.
\end{enumerate}
}

\section{Evaluation}
\label{sec:evaluation} \vspace{-0.2cm} The goal of this section is
to provide empirical evidence supporting the hypotheses postulated
in section~\ref{sec:introduction}. We show that the proposed
algorithm can help enrich the topic vocabulary and that the learned
vocabulary allows to generate queries that result in better
retrieval performance than queries generated directly from the
initial vocabulary.

To perform our tests we used nearly 500 topics from the Open
Directory Project (ODP)\footnote{http://dmoz.org}. The topics were
selected from the third level of the ODP hierarchy. A number of
constraints were imposed on this selection with the purpose of
ensuring the quality of our test set. The minimum size for each
selected topic was 100 URLs and the language was restricted to
English. For each topic we collected all of its URLs as well as
those in its subtopics. The total number of collected pages was more
than 350000. The Terrier framework~\cite{ounis07terrier} was used to
index these pages and to run our experiments.

In our tests we used the ODP description of each selected topic to
create an initial context description $\mathcal{C}$. The proposed
algorithm was run for each topic for at least $v=100$ iterations,
with 10 queries per iteration and retrieving 10 results per queries.
To create the queries $\mathcal{Q}$ at each iteration we used the
roulette selection mechanism. Roulette selection is a technique
typically used by Genetic Algorithms~\cite{holland75adaptation} to
choose potentially useful solutions for recombination, where the
fitness level is used to associate a probability of selection. In
our case, the fitness level was determined by the descriptive or
discriminating power values of the terms. The descriptor and
discriminator lists were limited to up to 100 terms each. The other
parameters in our algorithm were set as follows: $u=10$,
$\alpha$=0.5, $\beta$=0.5, $\gamma$=0.33, $\zeta$=0.33, $\xi$=0.33,
$\mu$=0.2 and $\nu$=0.1. In addition, we used the stopword list
provided by Terrier, Porter stemming was performed on all terms and
none of the query expansion methods offered by Terrier was applied.

To analyze the evolution of the context vocabulary we propose here a
revised notion of similarity. This measure of similarity is based on
$\simi$ but disregards the terms that form the query, favoring the
exploration of new material. Given a set of queries
$\{q_0,\ldots,q_p\}$ we define a novelty-driven similarity measure
$\nsimilarity: \{q_0,\ldots,q_p\} \times \{d_0,\ldots, d_{m-1}\}
\times \{d_0,\ldots, d_{m-1}\} \rightarrow [0,1]$ as: {\small
\[\nsimilarity({\bf q},d_i,d_j)= \simi(d_i - {\bf q}, d_j - {\bf
q})
\]}
The notation $d_i-{\bf q}$  stands for the representation of the
document $d_i$  with all the values corresponding to the terms from
query ${\bf q}$ set to zero. The same applies to $d_j-{\bf q}$.

\begin{center}
\begin{figure}[!ht]
\ \ \includegraphics[scale=0.35]{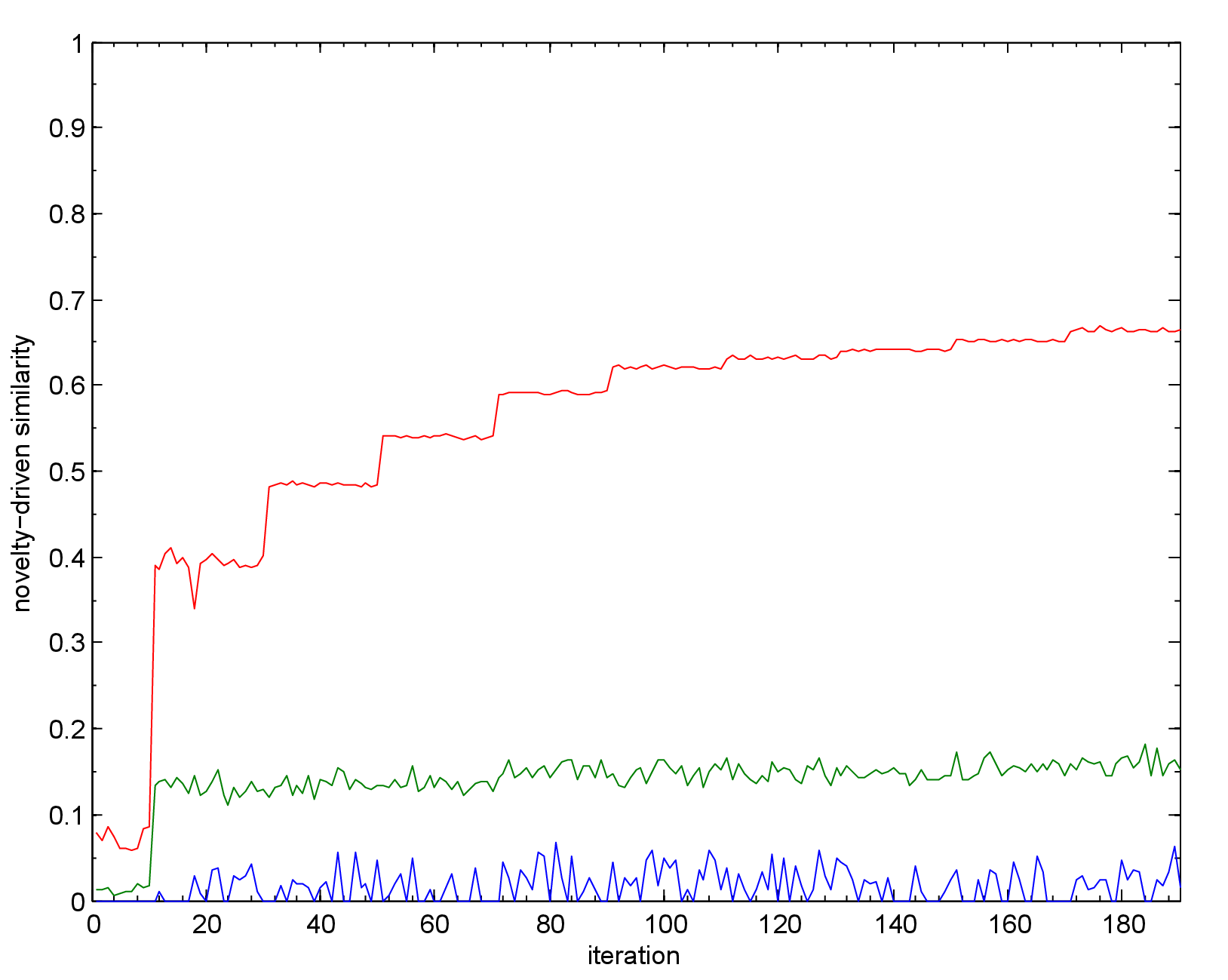}
\includegraphics[scale=0.36]{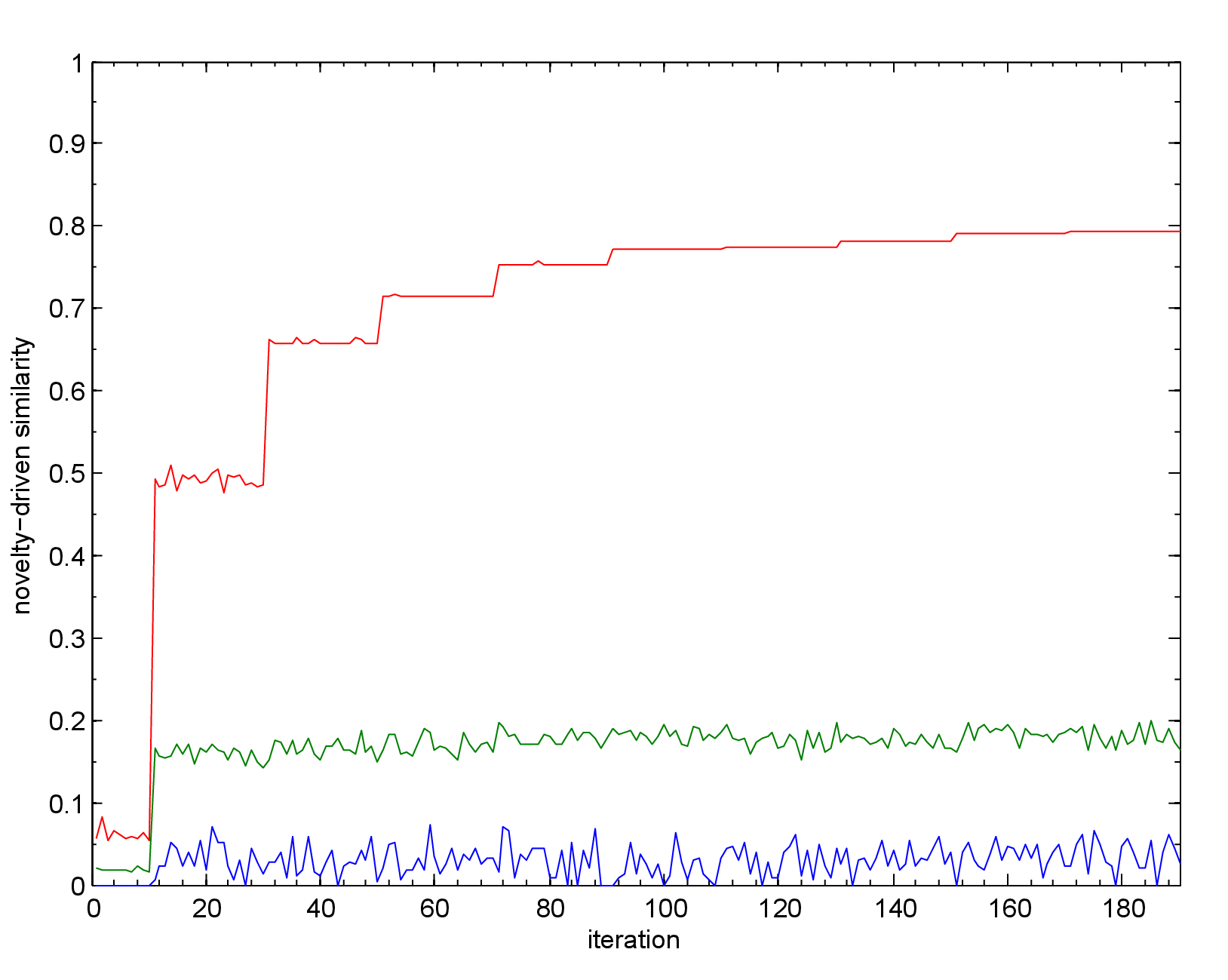} \caption{The evolution of minimum,
average and maximum novelty-driven similarities for the topics
{\tt\small Top$/$Home$/$Cooking$/$For\_Children} (left) and
{\tt\small Top$/$Computers$/$Open\_Source$/$Software} (right).}
\label{fig:evolution-novel-sim}
\end{figure}
\end{center}

\vspace{-0.7cm} We computed the novelty-driven similarity measure
$\nsimilarity$ between the initial context (topic descriptions) and the
retrieved results. The goal was to investigate the impact that each phase
change had on the query performance. Figure~\ref{fig:evolution-novel-sim}
shows the evolution of the novelty-driven similarity for the topics {\tt\small
Top$/$Home$/$Cooking$/$For\_Children} and {\tt\small
Top$/$Computers$/$Open\_Source$/$Software}.\footnote{For figures showing the
evolution of the novelty-driven similarity for each analyzed topic visit
\url{http://cs.uns.edu.ar/~cml/group/SimsCLEI08.htm}} We used the minimum,
average and maximum novelty-driven similarity between the initial context and
the search results at each iteration to illustrate the evolution of the
context vocabulary. It is worth noticing that the vocabulary leaps that
generally take effect every 10 iterations have an important impact on the
quality of the retrieved material. This provides evidence that the proposed
algorithm can help enrich the topic vocabulary. \vspace{-0.7cm}
\begin{center}
\begin{figure}[!ht]
\ \ \includegraphics[scale=0.32]{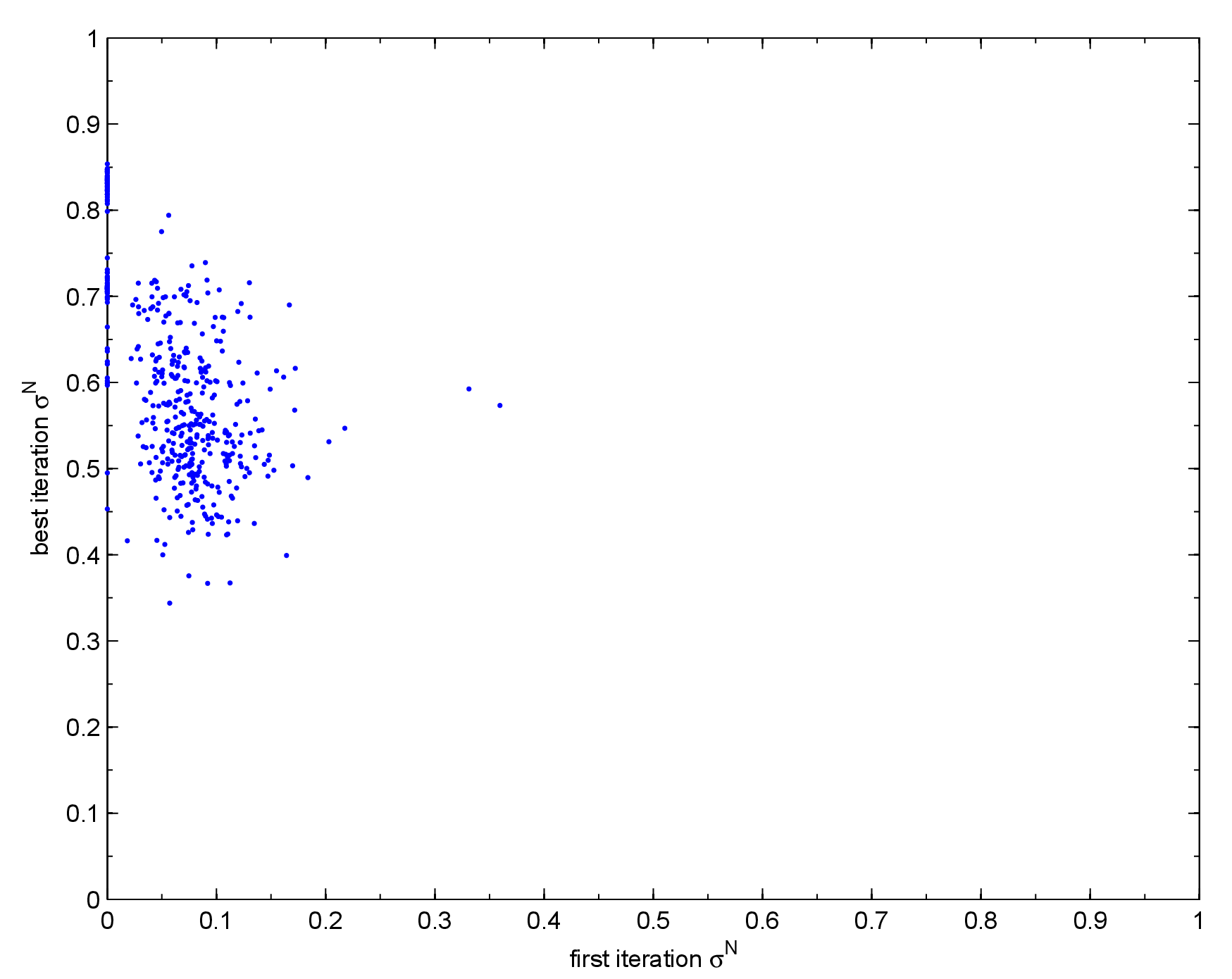}
\includegraphics[scale=0.32]{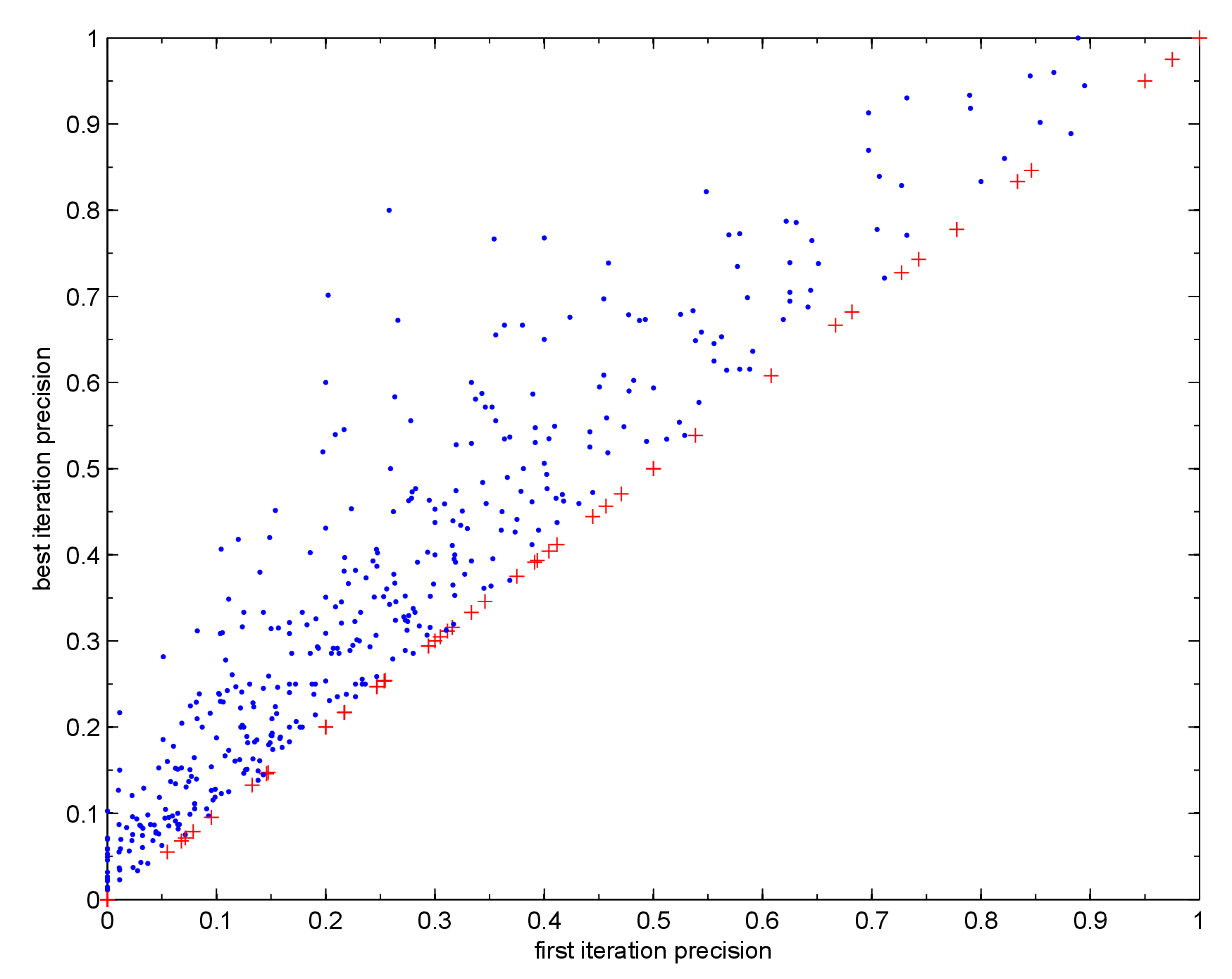}

\ \ \includegraphics[scale=0.32]{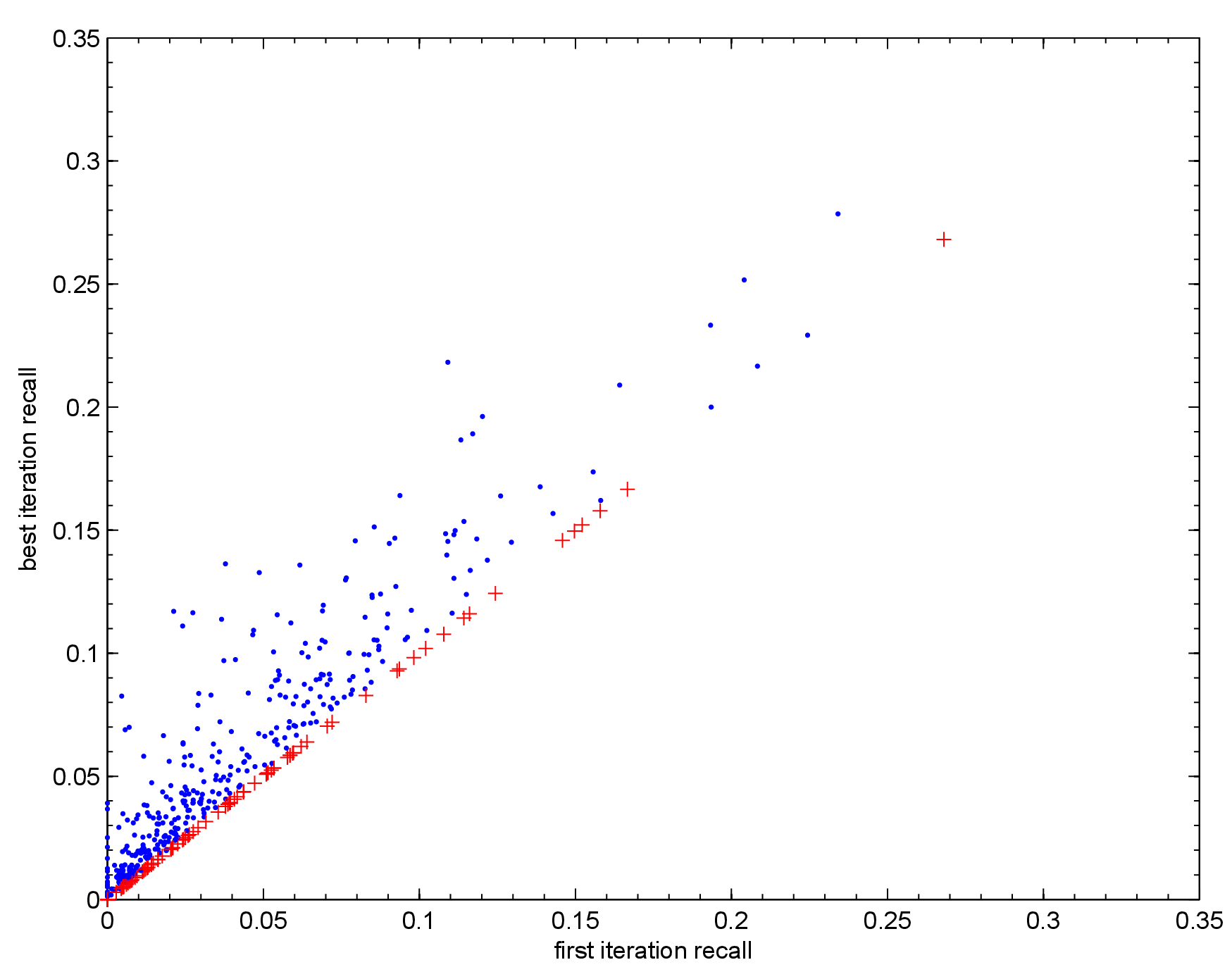}
\includegraphics[scale=0.32]{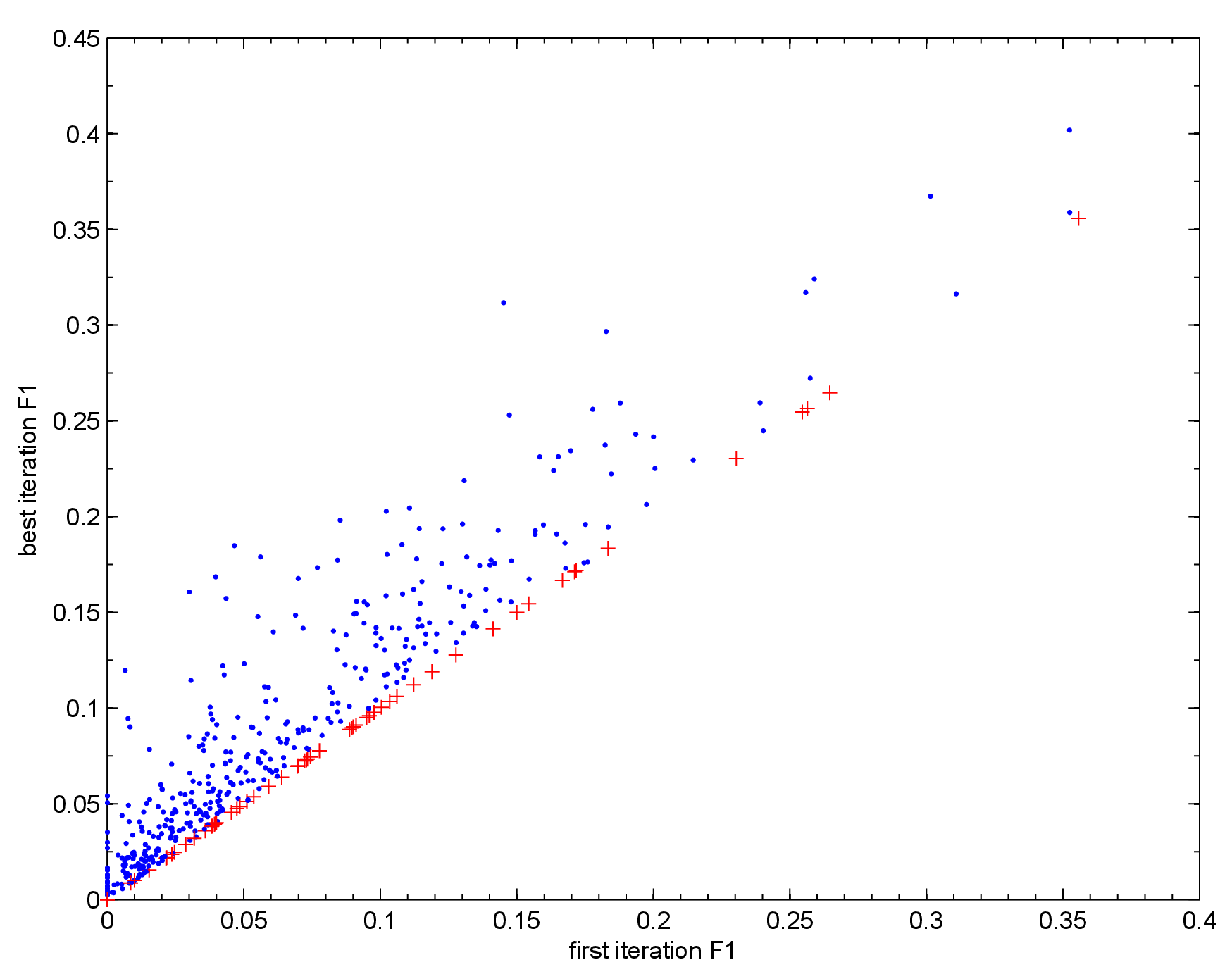}
\caption{Comparison of query performance for the first iteration vs.
the best iteration.} \label{fig:novelty-precison-recall-f1score}
\end{figure}
\end{center}
\vspace{-1cm}
 After observing that our algorithm had an impact on
the retrieval performance, our next step was to quantify this
impact. With that purpose we computed four measures of query quality
for the queries formed using the initial vocabulary and for the
queries constructed using the evolved vocabulary. The measures used
for this performance comparison are (1) maximum novelty-driven
similarity, (2) precision (fraction of retrieved documents which are
known to be relevant), (3) recall (fraction of known relevant
documents which were effectively retrieved), and (4) the harmonic
mean F1 (a measure which combines recall and precision). For a
detailed description of these well-known performance metrics we
refer the reader to any IR textbook
(e.g.,~\cite{baeza-yates99modern}). It is worth mention that the
relevant set for each analyzed topic was set as the collection of
its URLs as well as those in its subtopics.

The charts in figure~\ref{fig:novelty-precison-recall-f1score}
compare the performance of queries using the initial vocabulary
against queries using the evolved vocabulary. Each of the topics
corresponds to a trial and is represented by a point. The point's
horizontal coordinate corresponds to the performance of the queries
at the first iteration (initial vocabulary), while the vertical
coordinate corresponds to the performance of the queries at the best
iteration (evolved vocabulary). The points above the diagonal
corresponds to cases in which an improvement is observed for the
evolved vocabulary. In this evaluation, queries constructed using
the evolved vocabulary outperform the initial ones in 100\% of the
cases for novelty-driven similarity, 89.18\% of the cases for
precision, 89.38\% of the cases for recall, and 89.38\% of the cases
for the harmonic mean F1. It is interesting to note that for all the
topics analyzed the system managed to identify a better context
characterization as evidenced by the 100\% improvement for the
novelty-driven similarity performance metric. This highlights the
usefulness of evolving the context vocabularies to discover good
query terms.

Novelty-driven similarity and precision are useful metrics at the
time of evaluating the performance of IR systems that recover a few
pages out of a large set of relevant documents. This is the case for
our particular scenario and therefore we can use these two metrics
to statistically analyze the improvements achieved by the proposed
algorithm. In table~\ref{table:statistics} we present the means and
confidence intervals resulting from this analysis. These comparison
tables show that the use of an evolved vocabulary results in
statistically significant improvements over the use of the initial
vocabulary. \vspace{-0.7cm}
\begin{center}
\begin{table}[!ht]
\ \ \ \ \begin{tabular}{|l|c|c|c|}
  \hline
  $\nsimilarity$ & N & Mean & 95\% CI \\
  \hline
  first iteration    & 449 & 0.0661 & [0.0618;0.0704]  \\
  best iteration & 449 & 0.5970 & [0.5866;0.6073]  \\
  \hline
\end{tabular} \ \ \ \ \ \ \ \
\begin{tabular}{|l|c|c|c|}
  \hline
  Precision & N & Mean & 95\% CI\\
  \hline
  first iteration    & 449 & 0.2662 & [0.2461;0.2863]\\
  best iteration & 449 & 0.3538 & [0.3318;0.3757]\\
  \hline \end{tabular}

\vspace{0.3cm}
 \caption{Statistical analysis comparing query
performance for the initial vocabulary (first iteration) vs. query
performance for the evolved vocabulary (best iteration).}
\label{table:statistics}
\end{table}
\end{center}
\vspace{-2cm}
\section{Related Work}
\label{sec:related} \vspace{-0.2cm}

Extensions to basic IR approaches have examined some of the issues
raised in this paper. For instance, some automatic relevance
feedback techniques, such as the Rocchio's method
\cite{rocchio71relevance}, make use of the full search context for
query refinement. In these approaches the original query is expanded
by adding a weighted sum of terms corresponding to relevant
documents, and subtracting a weighted sum of terms from irrelevant
documents. As a consequence the terms that occur often in documents
similar to the input topic will be assigned the highest rank, as in
our descriptors. However, our technique also gives priority to terms
that {\em occur only in relevant documents} and not just to those
that {\em occur often}. In other words, we prioritize terms for both
discriminating and descriptive power. The  techniques for query term
selection proposed in this paper share insights and motivations with
other methods for query expansion and
refinement~\cite{scholer02query,billerbeck03query}. However, systems
applying these methods differ from our framework in that they
support this process through a query or browsing interfaces
requiring explicit user intervention, rather than formulating
queries automatically.

Our techniques rely on the notions of document similarity to
discover higher-order relationships in collections of documents.
This relates to the use of LSA~\cite{deerwester90indexing} to
uncover the latent relationships between words in a collection. Less
computationally expensive techniques are based on mapping documents
to a kernel space where documents that do not share any term can
still be close to each other~\cite{cristianini02latent}. Another
corpus-based technique that has been applied to estimate semantic
similarity is PMI-IR~\cite{turney01mining}, which measures the
strength of association between two elements (e.g., terms) by
contrasting their observed frequency against their expected
frequency. Differently from our proposal, the goal of these
techniques is to estimate the semantic distance between terms and
documents, without identifying topic descriptors and discriminators.

\vspace{-0.2cm}
\section{Conclusions}
\label{sec:conclusions} \vspace{-0.1cm}

In this paper we have presented an intelligent IR approach for
learning context-specific terms. Based on this approach, an
intelligent system can take advantage of the information available
in the user context to perform  search on the Web or other
information retrieval systems. We have shown that the user context
can be usefully exploited to access relevant material. However,
terms that occur in the user context are not necessarily the most
useful ones. In light of this we have proposed an incremental method
for context refinement based on the analysis of search results. We
also distinguish two natural notions, namely topic descriptors and
topic discriminators. The proposed notions are useful for meaning
disambiguation and therefore can help deal with the problem of
polysemy. Our evaluations show the effectiveness of incremental
methods for learning better vocabularies and for generating better
queries.

 Learning better
vocabularies is a way to increase the awareness and accessibility of
useful material. We have proposed a promising method to identify the
need behind the query, which is one of the main goals for many
current and next generation Web services and tools. As part of our
future work we expect to investigate different parameter settings
for the proposed algorithm and to develop methods that automatically
learn and adjust these parameters. In addition, we expect to run
additional tests comparing our approach with other existing query
refinement mechanisms.


\begin{thebibliography}{10}

\bibitem{baeza-yates99modern}
Ricardo Baeza-Yates and Berthier Ribeiro-Neto.
\newblock {\em Modern Information Retrieval}.
\newblock Addison-Wesley, 1999.

\bibitem{balabanovic95adaptive}
Marko Balabanovic, Yoav Shoham, and Yeogirl Yun.
\newblock An adaptive agent for automated {Web} browsing.
\newblock {\em Journal of Visual Communication and Image Representation}, 6(4),
  1995.

\bibitem{billerbeck03query}
Bodo Billerbeck, Falk Scholer, Hugh~E. Williams, and Justin Zobel.
\newblock Query expansion using associated queries.
\newblock In {\em Proceedings of the twelfth international conference on
  Information and knowledge management}, pages 2--9. ACM Press, 2003.

\bibitem{budzik01information}
Jay Budzik, Kristian~J. Hammond, and Larry Birnbaum.
\newblock Information access in context.
\newblock {\em Knowledge based systems}, 14(1--2):37--53, 2001.

\bibitem{cristianini02latent}
Nello Cristianini, John Shawe-Taylor, and Huma Lodhi.
\newblock Latent semantic kernels.
\newblock {\em J. Intell. Inf. Syst.}, 18(2-3):127--152, 2002.

\bibitem{deerwester90indexing}
Scott~C. Deerwester, Susan~T. Dumais, Thomas~K. Landauer, George~W. Furnas, and
  Richard~A. Harshman.
\newblock Indexing by latent semantic analysis.
\newblock {\em Journal of the American Society of Information Science},
  41(6):391--407, 1990.

\bibitem{etzioni96moving}
Oren Etzioni.
\newblock Moving up the information food chain: Deploying {Softbots} on the
  {World Wide Web}.
\newblock In {\em Proceedings of the Thirteenth National Conference on
  Artificial Intelligence and the Eighth Innovative Applications of Artificial
  Intelligence Conference}, pages 1322--1326, Menlo Park, 4--8 1996. AAAI Press
  / MIT Press.

\bibitem{gordon88necessity}
M.~Gordon.
\newblock The necessity for adaptation in modified boolean document retrieval
  systems.
\newblock {\em Information Processing and Management}, 24(3):339--347.

\bibitem{holland75adaptation}
John~H. Holland.
\newblock {\em Adaptation in natural and artificial systems}.
\newblock Ann Arbor: The University of Michigan Press, 1975.

\bibitem{licklider60manmachine}
J.~C.~R. Licklider.
\newblock Man-machine symbiosis.
\newblock {\em IRE Transactions on Human Factors in Electronics}, HFE-1:4--11,
  March 1960.

\bibitem{lorenzetti07intelligent}
Carlos~M. Lorenzetti, Rocio~L. Cecchini, and Ana~G. Maguitman.
\newblock Intelligent methods for information access in context: The role of
  topic descriptors and discriminators.
\newblock In {\em Proceedings of VIII Workshop de Agentes y Sistemas
  Inteligentes - CACIC 2007: XIII Congreso Argentino de Ciencias de la
  Computaci\'on}, 2007.

\bibitem{maes94agents}
Pattie Maes.
\newblock Agents that reduce work and information overload.
\newblock {\em Communications of the ACM}, 37(7):30--40, 1994.

\bibitem{maguitman05suggesting}
Ana Maguitman, David Leake, and Thomas Reichherzer.
\newblock Suggesting novel but related topics: towards context-based support
  for knowledge model extension.
\newblock In {\em IUI '05: Proceedings of the 10th international conference on
  Intelligent user interfaces}, pages 207--214, New York, NY, USA, 2005. ACM
  Press.

\bibitem{maguitman04dynamic}
Ana Maguitman, David Leake, Thomas Reichherzer, and Filippo Menczer.
\newblock Dynamic extraction of topic descriptors and discriminators: Towards
  automatic context-based topic search.
\newblock In {\em Proceedings of the Thirteenth Conference on Information and
  Knowledge Management (CIKM)}, Washington, DC, November 2004. ACM Press.

\bibitem{ounis07terrier}
I.~Ounis, C.~Lioma, C.~Macdonald, and V.~Plachouras.
\newblock Research directions in {Terrier}.
\newblock {\em Novatica/UPGRADE Special Issue on Web Information Access,
  Ricardo Baeza-Yates et al. (Eds), Invited Paper}, 2007.

\bibitem{ramirez06semantic}
Eduardo~H. Ramirez and Ramon~F. Brena.
\newblock Semantic contexts in the internet.
\newblock In {\em LA-WEB '06: Proceedings of the Fourth Latin American Web
  Congress}, pages 74--81, Washington, DC, USA, 2006. IEEE Computer Society.

\bibitem{rhodes00justintime}
Bradley Rhodes and Pattie Maes.
\newblock Just-in-time information retrieval agents.
\newblock {\em IBM Systems Journal special issue on the MIT Media Laboratory},
  39(3-4):685--704, 2000.

\bibitem{rocchio71relevance}
J.~J. Rocchio.
\newblock Relevance feedback in information retrieval.
\newblock In G.~Salton, editor, {\em The Smart retrieval system - experiments
  in automatic document processing}, pages 313--323. Englewood Cliffs, NJ:
  Prentice-Hall, 1971.

\bibitem{scholer02query}
Falk Scholer and Hugh~E. Williams.
\newblock Query association for effective retrieval.
\newblock In {\em Proceedings of the eleventh international conference on
  Information and knowledge management}, pages 324--331. ACM Press, 2002.

\bibitem{turney01mining}
Peter~D. Turney.
\newblock Mining the web for synonyms: {PMI-IR} versus {LSA} on {TOEFL}.
\newblock In {\em EMCL '01: Proceedings of the 12th European Conference on
  Machine Learning}, pages 491--502, London, UK, 2001. Springer-Verlag.

\end{thebibliography}
 \end{document}